\documentclass{elsart}

\usepackage{graphicx}
\usepackage{epsfig}

\usepackage{amssymb}
\usepackage{wasysym}

\begin{document}
\begin{frontmatter} 

\title{Ambivalent effects of added layers on steady kinematic
dynamos in cylindrical geometry: application to the VKS
experiment}

\author[ROSS]{F. Stefani\corauthref{cor1}}
\author[ROSS]{, M. Xu}
\author[ROSS]{, G. Gerbeth}
\author[SACLAY]{, F. Ravelet}
\author[SACLAY]{A. Chiffaudel}
\author[SACLAY]{, F. Daviaud}
\author[MEUDON]{, J. L\'{e}orat}
\corauth[cor1]{Corresponding author. e-mail: F.Stefani@fz-rossendorf.de}
\address[ROSS]{Forschungszentrum Rossendorf, P.O. Box 510119,
D-01314 Dresden, Germany}
\address[SACLAY]{Service de Physique de l'{\'E}tat Condens\'{e}, DSM, CEA Saclay, CNRS URA 2464,
91191 Gif-sur-Yvette, France}
\address[MEUDON]{LUTH, Observatoire de Paris-Meudon, 92195 Meudon, France}

%\date{\today}

\begin{abstract}
The intention of the ''von K\'{a}rm\'{a}n sodium'' (VKS) experiment is to study
the hydromagnetic dynamo effect in a highly turbulent and unconstrained flow.
Much effort has been devoted to the optimization of the mean flow  and
the lateral boundary conditions in order to minimize the critical
magnetic Reynolds number $R^c_m$ and hence the necessary motor power.
The main focus of this paper lies on the role of ''lid layers'', i.e.
layers of liquid sodium between the impellers and the end walls of the cylinder.
First, we study an analytical test flow to show that
lid layers can have an ambivalent effect on the efficiency of the dynamo.
The critical magnetic Reynolds number shows a flat minimum for
a small lid layer thickness, but increases for thicker layers.
For the actual VKS geometry it is shown
that static lid layers yield a moderate increase
of $R^c_m$ by approximately 12 per cent. A more
dramatic increase by 100 till 150 per cent                                         %!!!
can occur when some rotational flow is taken into account in those layers.
Possible solutions of this problem are discussed for the real dynamo
facility.
\end{abstract}


\end{frontmatter}

\section{Introduction}

The Earth's magnetic field, as most other cosmic magnetic fields, 
is generated by the hydromagnetic dynamo effect. \cite{RUHO}.
The dynamo effect is an instability of
the magnetic field in a flow of
conducting liquid, which is  controlled by
the magnetic Reynolds
number $R_m=\mu \sigma L V$, where
$\mu$ is the magnetic permeability of the fluid
(which is, in most cases, equal to the
permeability of the vacuum, $\mu_0$),
$\sigma$ the electrical conductivity of the fluid, and
$L$ and $V$ are typical length and
velocity scales of the flow, respectively.
Typical values of the critical $R_m$ are in the order
10 till 100. A comparably simple way to reach such values 
is the use of
materials with high relative permeability. 
Indeed, this  was  done by Lowes and Wilkinson 
in the sixties. 
Their first successful homogeneous dynamos comprised 
two Perminvar cylinders   spinning 
around non-parallel axes
in a ''house-shaped'' surrounding
conductor \cite{LOWI1,LOWI2}. 
Although being genuine homogeneous dynamos, 
they did not allow to study non-trivial back-reaction
effects as they are typical for the {\it hydro}magnetic dynamos
in the cosmos.

The first experimental realisation of such hydromagnetic 
dynamos was left to 1999 \cite{RMP}, when 
magnetic field self-excitation was
observed in two large-size liquid sodium
facilities in Riga \cite{PRL1,PRL2} and Karlsruhe \cite{MUST,STMU}.
Since that time, both
experiments have brought about a wealth of
data on the kinematic and the saturated regime.
In both cases, the comparison of the experimental results
with the numerical predictions shows  a satisfactory
agreement.
In contrast to electro-technical dynamos which are characterized by
an abrupt saturation at the critical
rotation rate due to their constructional
stiffness, hydromagnetic dynamos saturate in a much
''softer'' way
due to the
deformability of the flow structure. This has been
shown, in particular,
at the
Riga experiment \cite{PLASMA}, where the back-reaction of
the magnetic field
acts very selectively on the azimuthal velocity
component. This flow deformation explains the very
flat increase of the Joule power above the
self-excitation threshold.

While both the kinematic dynamo
effect and the non-trivial saturation mechanism (due
to flow deformation) have been investigated in the Riga 
experiment,
there are still open questions
which are better addressed in the
framework of other experimental facilities.
One of the most important issues concerns the role of
high levels of turbulence on the self-excitation condition,
a problem that can hardly be solved
by theory and numerics alone.

\begin{figure}
\begin{center}
\begin{tabular}{c}
\epsfxsize=10.0cm\epsfbox{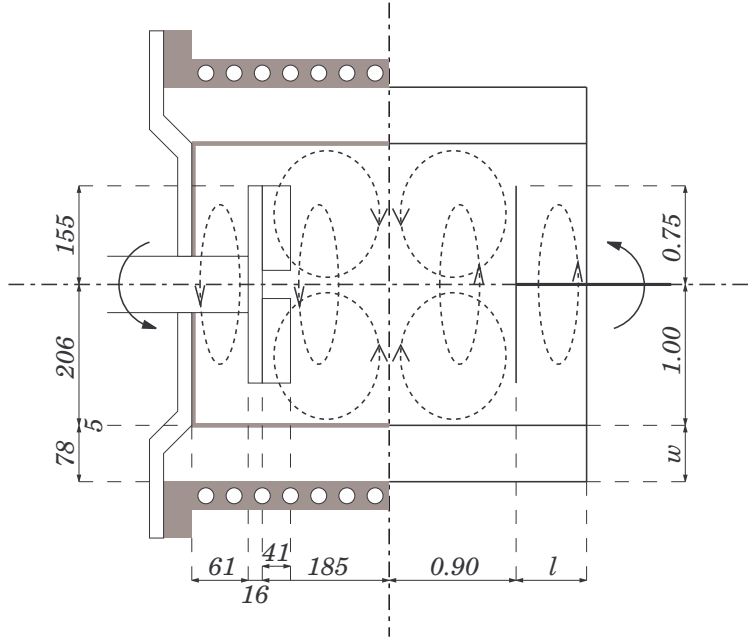}
\end{tabular}
\end{center}
\caption{Principle design, flow structure, and dimensions
of the VKS experiment. The two impellers produce
two counter-rotating toroidal eddies and, by
centrifugal pumping, two
poloidal rolls.
Left half: Technical details, including
the main vessel with the cooling system, the copper envelope,
the impeller with attached blades and its shaft.
Copper parts are gray, stainless steel parts are white.
The indicated
dimensions are in mm. Right half: Simplified geometry as
it is assumed in the numerical simulations with dimensionless
lengths. The dimensionless
thicknesses of the side layer and the lid layer are
denoted by $w$ and $l$, respectively. Note that in the VKS2
device, these values are close to $w=$ 0.415
(the presence of a copper inner wall may be modeled using
$w=$ 0.45)  and $l=$ 0.385, while $w$ and $l$ are taken as
parameters in the numerical computations.}
\end{figure}

Besides other experimental efforts at different sites
in the world \cite{RMP}, the VKS project in Cadarache, France,
is one of the most promising candidates to meet this
goal \cite{VKS1,VKS1.5,VKS2}.
The acronym ''VKS''  stands for ''von K\'arm\'an sodium''
and refers to a flow that is produced between two
counter-rotating impellers in a finite cylinder (cf. Fig. 1).
The
impellers consist of flat disks fitted with blades, which
ensures a very efficient inertial stirring. The phenomenology
of the mean flow is the following. Each impeller acts as a
centrifugal pump: the fluid rotates with the impeller and is
expelled radially by the centrifugal effect. To ensure mass
conservation the fluid is pumped in the center of the
impeller and recirculates near the cylinder wall. In the
exact counter-rotating regime, the mean flow is divided into
two toric cells separated by an azimuthal shear layer.
The integral Reynolds number for this experiment is of the
order of $10^7$ and the shear layer instability is a
strong source of turbulence.

In the first version of this experiment, called ''VKS1'',
the achieved magnetic Reynolds number $R_m$ was below the
numerical prediction for criticality, computed for the time-averaged
flow \cite{MBDL}, and no self-excitation of a magnetic field was
observed. Various measurements of induced magnetic field have
been performed \cite{VKS1,VKS1.5,VKS2}, showing a good agreement
with preceding numerical predictions. In a second version (VKS2),
which is still under commissioning,
a higher motor power (300 kW) was installed.
A careful optimization of the time-averaged flow has been
carried out, using a water-model experiment and varying the
design of the driving impellers \cite{LEORAT}. A solution
achievable in the experimental device VKS2 has been found.
This solution is based on the adjunction of a layer of
stationary conductor surrounding the flow, which was supposed to 
reduce the
critical magnetic Reynolds number $R_m^c$ by a
factor of 4. It will be called ''side layer'' and
its thickness will be denoted by $w$ in the following.

The kinematic dynamo simulations were carried out
with a code that
uses periodic boundary conditions in axial direction
\cite{LEORAT,MBDL},
what has to be
considered as a compromise. Of course, the correct
implementation of magnetic boundary conditions for
non-spherical bodies has been
a longstanding problem in dynamo research. Only recently,
new methods and codes have been proposed to overcome this problem
\cite{STEF,ASTRO,GUERMOND,JCP,ISKAKOV,PRE,BOUR}.

One of the goals of this paper is to apply two of
those recently established numerical schemes to 
the design of cylindrical fluid dynamos, such as 
the
VKS dynamo, and to
compare their results with
the results obtained with the periodic code.
These two codes will be shortly described in section 2.
The first one
is a
finite-difference scheme coupled to a solver of
the Laplace equation in the
external vacuum region, which had been successfully
used in the simulation of the Riga
dynamo experiment
\cite{PLASMA,STEF}.
The second one is a code based on the
integral equation approach
for dynamos in finite domains
\cite{ASTRO,JCP,PRE}
which basically relies on
the self-consistent treatment of Biot-Savart's law.

These two codes will first be applied to an analytical test
flow in section 3.
Interestingly, we will see that the results obtained
by the two codes do not differ
significantly from those obtained with the
periodic code.

Furthermore, these two non-periodic codes allow a more
realistic treatment of the VKS dynamo system.
We obtained surprising and unsettling results
when taking into account the geometry of the vessel 
in the axial direction behind the impellers (section 4).
This is characterized by the
existence of layers of liquid sodium between the
impellers and the end walls (the ''lids'') of the cylindrical
vessel (Fig. 1). In the following, these layers will be
referred  as "lid layers", in distinction from the ''side layers''
at the envelope of the cylinder. 
We will see that even the mere existence of
static lid layers
increases
the critical magnetic Reynolds number $R^c_m$ by some 12
per cent.
This is already a remarkable result as it shows,
contrary to common wisdom
\cite{BUGU,AVALOS2}, that  {\it even for
non-oscillatory} eigenfields the existence of conducting layers
is not
always advantageous for the dynamo to work
(for {\it oscillatory}
eigenfields, the possible deterioration of the dynamo condition
with increasing layer thickness
was discussed in \cite{TILGNER,AVALOS}).
The most dramatic effect occurs, however,
when some rotational flow
is assumed in the lid layer, which must be considered as
naturally
driven by the impeller. Then, even values of $R^c_m$ close to  130
can be found
depending on the details of the assumed flow there.

The paper closes with a discussion of the results and
with proposals how to overcome
the indicated problems by constructional add-ons.

\section{The utilized numerical codes}

As stated in the first section, the periodic magnetic boundary 
conditions are easy to implement and the simplicity of the resulting 
kinematic dynamo code allows parametric studies without involving much 
numerical resources.
This pseudo-spectral code with {\it periodic}
boundary conditions for the magnetic field \cite{LEORAT0,LEORAT,MBDL}, which uses the
Adams-Bashforth method for the time-stepping, will be
denoted  by ''PER'' in the following.
As the real experiments take place obviously in a
finite cylindrical container, it is particularly interesting  to 
numerically verify if $R^c_m$ for a flow 
configuration which is optimal in the periodic case is  a robust 
quantity when the magnetic boundaries become finite. For this purpose, 
we have employed two different codes. 
The first
one, henceforth referred as the
{\it differential equation approach} (DEA),
relies on the induction equation
\begin{eqnarray}
\frac{\partial {{\bf{B}}}}{\partial t}=\nabla
\times ({\bf{u}} \times {\bf{B}})
+\frac{1}{\mu_0 \sigma} \Delta {\bf{B}}\; ,
\end{eqnarray}
wherein ${\bf{B}}$ is the magnetic field and ${\bf{u}}$
the velocity field.
In this paper we will restrict our attention to
axisymmetric flows which allows a decoupled treatment
of the
azimuthal modes $\sim \exp{(i m\varphi)}$ of the magnetic field.
Focusing on the azimuthal
mode with $m=1$, which is known
to be the dominant mode for the VKS dynamo, the DEA code
is a finite difference solver
with a non-uniform grid in radial and axial direction.
The time evolution is realized as an Adams-Bashforth scheme.
The
vacuum boundary
conditions are
implemented in such a way that for each time step the
Laplace equation in the outer part is solved
(by pseudo-relaxation),
whereafter the resulting external
solution is matched to the inner solution via the
continuity demands
for the tangential electric field components and
the divergence free
condition for
the magnetic field. This code was already used
successfully
for the
prediction and optimization of the Riga dynamo
experiment \cite{PLASMA,STEF}.

The second code is based on the {\it integral equation
approach} (IEA) to dynamos in finite domains,
which had been developed
and qualified for real problems
during the last years \cite{ASTRO,JCP,PRE}.
Basically, it uses the three coupled integral equations
\begin{eqnarray}
{\mathbf{B}}({\mathbf{r}})&=&\frac{\mu_0\sigma}{4\pi}\iiint\limits_V \frac{({\mathbf{u}}
({\mathbf{r}}')\times{\mathbf{B}}({\mathbf{r}}'))\times({\mathbf{r}}-{\mathbf{r}}')}
{|{\mathbf{r}}-{\mathbf{r}}'|^3}dV' \nonumber \\
&&-\frac{\mu_0\sigma}{4\pi}\oiint\limits_S \phi({\mathbf{s}}'){\mathbf{n}}({\mathbf{s}}')
\times\frac{{\mathbf{r}}-{\mathbf{s}}'}{|{\mathbf{r}}-{\mathbf{s}}'|^3}dS'\nonumber \\
&&-\frac{\mu_0\sigma\lambda}{4\pi}\iiint\limits_V \frac{{\mathbf{A}}
({\mathbf{r}}')\times({\mathbf{r}}-{\mathbf{r}}')}
{|{\mathbf{r}}-{\mathbf{r}}'|^3}dV'\; ,\\
\frac{1}{2}\phi({\mathbf{s}})&=&\frac{1}{4\pi}\iiint\limits_V \frac{({\mathbf{u}}({\mathbf{r}}')
\times{\mathbf{B}}({\mathbf{r}}'))\cdot({\mathbf{s}}-{\mathbf{r}}')}{|{\mathbf{s}}-{\mathbf{r}}'|^3}dV' \nonumber \\
&&-\frac{1}{4\pi}\oiint\limits_S \phi({\mathbf{s}}'){\mathbf{n}}({\mathbf{s}}')\cdot
\frac{{\mathbf{s}}-{\mathbf{s}}'}{|{\mathbf{s}}-{\mathbf{s}}'|^3}dS'\nonumber \\
&&-\frac{\lambda}{4\pi}\iiint\limits_V \frac{{\mathbf{A}}({\mathbf{r}}')\cdot({\mathbf{s}}-{\mathbf{r}}')}
{|{\mathbf{s}}-{\mathbf{r}}'|^3}dV' \; ,\\
{\mathbf{A}}({\mathbf{r}})&=&\frac{1}{4\pi}\iiint\limits_V \frac{{\mathbf{B}}({\mathbf{r}}')
\times({\mathbf{r}}-{\mathbf{r}}')}{|{\mathbf{r}}-{\mathbf{r}}'|^3}dV'\nonumber\\
&&+\frac{1}{4\pi}\oiint\limits_S{\mathbf{n}}({\mathbf{s}}')\times\frac{{\mathbf{B}}({\mathbf{s}}')}{|{\mathbf{r}}-
{\mathbf{s}}'|}dS' \; .
\end{eqnarray}
The magnetic field $\bf B$ is determined by the volume integral equation
(2) which is
a rewritten form of Biot-Savart's law. The
vacuum boundary conditions are ensured by
the surface integral equation (3)
for the electric potential $\phi$, which
results from
applying Green's theorem to the solution of the
Poisson equation.
Equations (2) and (3), with the third terms on the r.h.s. omitted,
are sufficient for treating steady dynamo problems.
In the unsteady case, for which we
assume a time dependence $\sim \exp(\lambda t)$ for
all electromagnetic quantities,
they have to be completed
by the additional equation (4) for the vector potential $\bf A$.

Accordingly, we have implemented two versions of the IEA
approach. The
first one \cite{JCP}, using only Eqs. (2) and (3),
is an integral equation eigenvalue solver for
$R^c_m$ which works properly
only if the critical eigenmode is non-oscillatory, otherwise
it yields unphysical solutions with complex $R^c_m$.
The second version \cite{PRE}
is an integral equation
eigenvalue
solver
for the complex time constant
$\lambda$,
whose
real part is the growth rate, and whose
imaginary part is the angular frequency of the
magnetic eigenmode.

As in the DEA case, the magnetic field ${\mathbf{B}}$,
the electric potential $\phi$ and the vector
potential ${\mathbf{A}}$ are expanded in
azimuthal modes  $\sim \exp{(i m \varphi)}$.
The reduction of the equation system (2-4)
to the dominant mode with $m=1$ is
complicated and will be published elsewhere.
A few more remarks on this reduction can be found in the appendix.

\section{An analytical test example}

For the sake of comparison with existing results, and to
get a first feeling on the
role of lid layers, we have carried out some computations
for an analytical test flow that was proposed and treated in
\cite{MND,LEORAT}.
The topology of this flow, denoted by s2t2 \cite{DUJA},
is the same as in the real VKS experiment, consisting of two
counter-rotating toroidal eddies (t2) and two poloidal
rolls (s2).
The flow is defined by the following velocity field
in a cylindrical volume with $0 \le r \le 1$ and
$-1 \le z \le 1$:
\begin{eqnarray}
v_r&=&-\frac{\pi}{2} r (1-r)^2 (1+2r) \cos(\pi z)\; ,\\
v_{\varphi}&=&4 \epsilon r (1-r) \sin(\pi z/2)\; ,\\
v_z&=&(1-r)(1+r-5 r^2) \sin(\pi z)\; .
\end{eqnarray}

For the parameter $\epsilon$ which determines the ratio of
toroidal to poloidal flow, we have used the same value
$\epsilon=0.7259$
as in \cite{LEORAT}.
It is important to note that at the
envelope of the cylinder all three components
$v_r$, $v_{\varphi}$ and $v_z$
vanish while  at the top and the bottom
$v_r$ and $v_{\varphi}$ do not vanish.
Usually, the effect of any conducting layer around a dynamo
depends strongly on the existence or non-existence
of velocity components at the boundary.

Hereafter, we will always use the following definition of
the magnetic Reynolds number:
\begin{eqnarray}
R_m=\mu_0 \sigma r_1 v_{max}
\end{eqnarray}
where $r_1$ is the radius of the inner cylinder and
$v_{max}$ is the maximum velocity of the flow.

For the flow field according to Eqs. (5-7),
Ravelet et al. found values of
$R^c_m=58$
for side layer
thickness $w=0$, and $R^c_m=43$
for $w=1$ \cite{LEORAT}. These results were also
confirmed by an independent Galerkin code \cite{MND}.

\begin{figure}
\begin{center}
\begin{tabular}{c}
\epsfxsize=8.0cm\epsfbox{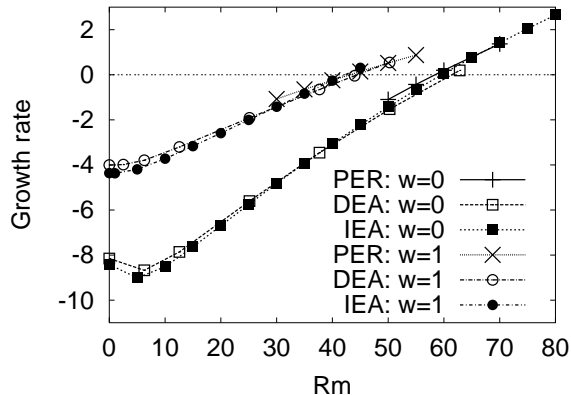}
\end{tabular}
\end{center}
\caption{Growth rates for the analytical test flow, with
side layer
thicknesses $w=0$ and $w=1$, as computed
by the PER, DEA, and IEA code.}
\end{figure}

In the DEA code, we have used a total of 119 grid points
in $z$ direction and 50 grid points
in $r$ direction. 65 of the grid points in $z$ direction are in the
internal part, the remaining 54 are outside.
For $w=0$, 21 of the grid points in
$r$ direction are in the internal part, the remaining  29 are outside.
For $w=1$, 39 of the grid points in
$r$ direction are in the internal part,
the remaining  11 are outside.
In the IEA code with $w=0$, we have used a 20 x 20 grid in $r$ and $z$
direction. For $w=1$, we have used 26 grid points in $r$ direction and
18 grid points in $z$ direction.

In Fig.  2 we show the computed growth rates for the
cases $w=0$ and $w=1$.
The critical values for all three cases are summarized in Table 1.
For $w=0$, the DEA code gives
$R^c_m=61.5$, the IEA code gives $R^c_m=59.6$.
For $w=0$, the DEA code gives
$R^c_m=44.5$, the IEA code gives $R^c_m=42.3$.
Interestingly enough, all three codes (PER, DEA, and IEA) give
comparable results, which are $R^c_m=59.8 \pm 1.8$ for $w=0$ and
$R^c_m=43.4 \pm 1.1$ for $w=1$.

\begin{table}
\caption{$R^c_m$ as computed for the analytical test flow with
two side layer thicknesses $w$ by
three different  codes explained in the text.}
\vspace*{0.5cm}
\begin{center}
\begin{tabular}{|l|l|l|}
\hline
Code&$w$&$R^c_m$\\
\hline
PER&0.0&58\\
DEA&0.0&61.5\\
IEA&0.0&59.6\\
\hline
PER&1.0&43\\
DEA&1.0&44.5\\
IEA&1.0&42.3\\
\hline
\end{tabular}
\end{center}
\end{table}

The following conclusions can be drawn:
For the considered analytical flow of the  s2t2 type,
the results of the DEA and the IEA code coincide quite
remarkably, despite
some differences in the used grid spacing.
What is more, our results
for $R^c_m$ differ only slightly from
the results that were obtained before by the PER code.
This agreement may be related to the fact
that the mirror-antisymmetry of the
magnetic eigenfield within the enlarged elementary cell
of the PER code ensures that no currents can leave the
flow region \cite{LEORAT}, which is indeed a
necessary boundary condition for the finite cylinder.
Although this correspondence might be a bit special
for the considered s2t2 flow, it
is nevertheless remarkable.
Its simplicity  
and reproducible dynamo
behaviour might qualify the analytical flow (5-7) 
as a benchmark for other kinematic dynamo codes in cylindrical 
geometry.

\begin{figure}
\begin{center}
\begin{tabular}{c}
\epsfxsize=8.0cm\epsfbox{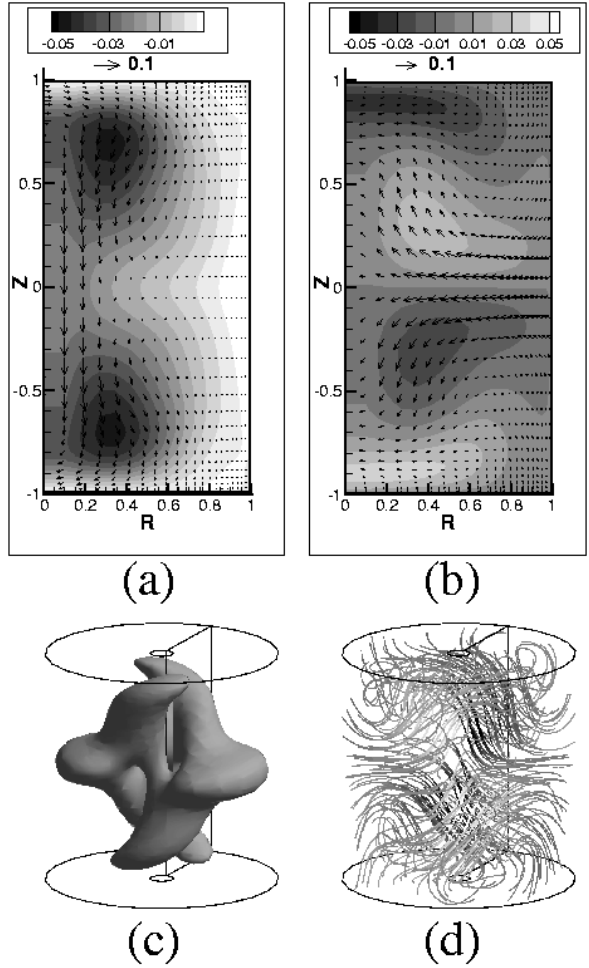}
\end{tabular}
\end{center}
\caption{(a,b) Poloidal field (arrows) and toroidal field (gray scale)
in two azimuthal planes  (rotated by 90 degrees),
computed with the DEA code for the analytical test flow.
(c) Isosurface of the magnetic field energy (25 per cent
of maximum). (d) Magnetic field lines, computed
with  the IEA code. The gray scale indicates the axial
component of the field. Note that the rotation axis
is vertical, contrary  to Fig. 1 and to the real VKS
experiment where it is horizontal.}
\end{figure}

For the case $w=0$, the structure of the magnetic
eigenfields is illustrated in Fig. 3. We show the poloidal
and toroidal field
structure in two azimuthal planes (Fig. 3a and 3b), which are
rotated by 90 degree with
respect to each other.
Figures 3c  and 3d
give impressions of the 3D structure of the fields.
In Fig. 3c we exhibit the isosurface
(25 per cent of the maximum value)
of the magnetic field energy,
Fig. 3d gives an impression of the magnetic field lines.
This eigenfield in form of an equatorial dipole
was already discussed in \cite{LEORAT} where also
the structure of the electric
currents is illustrated.

\begin{figure}
\begin{center}
\begin{tabular}{c}
\epsfxsize=8.0cm\epsfbox{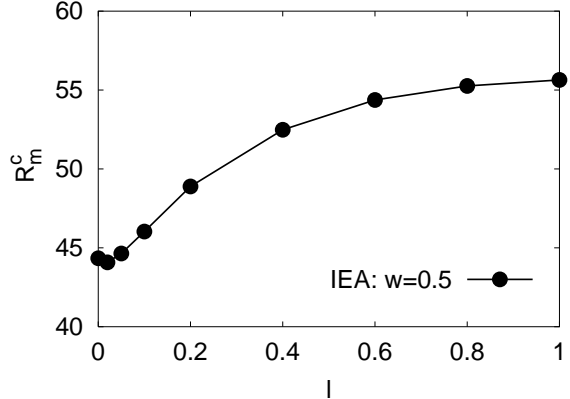}
\end{tabular}
\end{center}
\caption{Dependence of $R^c_m$ on the thickness $l$ of a stagnant lid layer 
for the
analytical test flow with fixed side layer thickness $w=0.5$.}
\end{figure}

Let us consider now the influence of a static lid layer
on $R^c_m$, which is documented in Fig. 4 for the IEA code.
We observe first a slight decrease of $R^c_m$ only
for very small lid layer thicknesses $l$,
while for larger $l$ the dynamo condition
deteriorates. For $l=1$ we get an increase of $R^c_m$
by approximately 20 per cent, compared to the value at $l=0$.
Note that the PER code, which uses Fourier transforms in
axial direction, is not very convenient for the handling of
sharp velocity gradients. Nevertheless, a few numerical tests 
with this code have qualitatively  confirmed the general 
tendencies observed with the DEA and IEA code.

This negative influence of lid layers on the
dynamo is quite in contrast
to the well-known positive effect of side layers
\cite{BUGU,AVALOS2}. Since such lid layers are indeed
present for technical reasons in the VKS2 experiment,
it is interesting to examine their role in more detail.

\section{The VKS flow}

After having considered  an analytical test flow,
we switch over now to  the flow of the real
VKS experiment. We have used
the time-averaged flow field of the propeller ''TM73'' which was
identified as the optimal flow
field in \cite{LEORAT}, according to the periodic code. 
After making some interpolations
to project this flow field onto the grids used in our
DEA and IEA codes, we have investigated the influence of side layers
and, in particular, the influence
of lid layers and a possible flow therein on the dynamo
condition.

In Table 2 we present a summary of  the  different lid layer and flow
settings that were
considered numerically, and
indicate the resulting  $R^c_m$. As before,
DEA and IEA refer to the codes
based on the differential equation approach and on the integral
equation
approach. $w$ is  the thickness of the side layers, $l$ the
thickness of the
lid layer. For numerical reasons, in some configurations
a certain smoothing of
the velocity field in
transition regions has been employed.
Probably, this will add to the
differences between the results of DEA and IEA.

\begin{figure}
\begin{center}
\begin{tabular}{c}
\epsfxsize=8.0cm\epsfbox{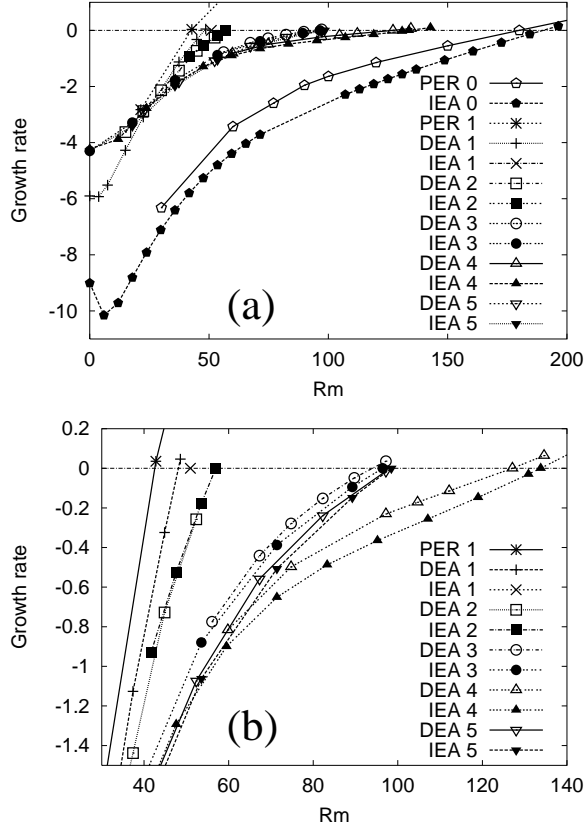}
\end{tabular}
\end{center}
\caption{Growth rates for the flow in the VKS experiment,
as computed with the DEA and the
IEA code for different settings of the lid layers and
the flow structure
therein. (b) is a zoom of (a). The results for PER 0 and PER 1 are also included.
See Table 2 for further details.}
\end{figure}

In Fig. 5, we have compiled the main results. Both parts
of the figure show mainly the same growth rate curves,
with more details visible in Fig. 5b.

The first observation concerns the very high
value $R^c_m=190$ for IEA 0,
which represents the case that neither a lid layer
nor a side layer is
present. A similar value
of 180 had been obtained with the periodic
code \cite{LEORAT}, and the corresponding growth rate curve
is indicated as PER 0 in Fig 5a.
By adding a side layer the DEA 1 and IEA 1 values drop to approximately
50 which is not far from the PER 1 result 43.
We see that now the effect of the side layer on
$R^c_m$ is much more
significant  than for the analytical test flow.
The reason for this difference between both gains was already
discussed in \cite{LEORAT}.

\begin{figure}
\begin{center}
\begin{tabular}{c}
\epsfxsize=8.0cm\epsfbox{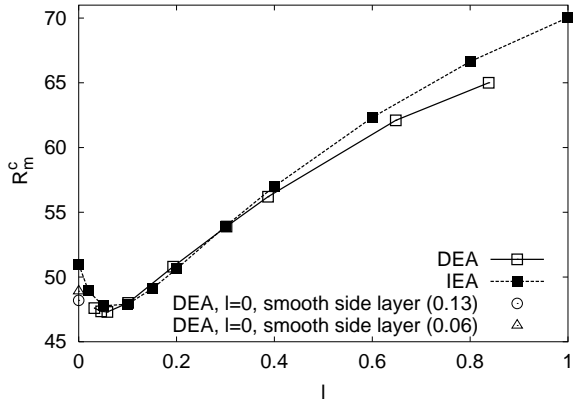}
\end{tabular}
\end{center}
\caption{Growth rates for the flow in the VKS experiment, as
computed with the DEA and the
IEA code for different thicknesses $l$ of a stagnant
lid layer.
At $l=0.4$ we obtain the cases DEA 2 and IEA 2 described
in Table 2.
The two individual points for DEA at $l=0$
correspond to a smoothing of the velocity field between
$r=1$ and $r=1.13$ or $r=1.06$, respectively.}
\end{figure}

Assume now the side layer thickness to be  fixed to $w=0.45$, and
consider the
dependence of $R^c_m$ on the lid layer thickness $l$,
first without any flow therein.
This dependence is given in Fig. 6.

Apart from a technical detail for the DEA code which
requires, for $l=0$, some
smoothing of the velocity at $r=1$, the IEA and DEA curves
are in
good correspondence,
with slightly larger deviations for higher values of $l$.
Again we observe a minimum of $R^c_m$
close to $l=0.05$,
indicating that a thin lid layer of liquid sodium
behind the propeller might be
advantageous for the dynamo.
However, with further increasing
$l$ this benefit is lost again, and for the (nearly)
real value $l=0.4$ we get $R^c_m=56$ (DEA 2) and
$R^c_m=57$ (IEA 2) 
which is approximately 12 per cent above the value for
$l=0$. This seems to be not very much, but in terms of the
necessary power consumption (which scales like $\sim R^3_m$)
it would amount to an increase of 40 per cent.

This negative effect becomes much more dramatic when we
consider the possibility of some flow in the lid layers.
It is
reasonable to
assume at least some rotation in these
layers due to the viscous coupling with the
impeller.
Up to now, this flow in the lid layers has not
been measured, and
it was not considered in the previous simulations with the
periodic code, which is probably  less suited to
handle steep velocity gradients.

We study here two simplified flow patterns in the lid layers which are,
at least, 
not completely unrealistic.

The first flow is basically a rotation which is linearly decaying
from the impeller positions at $z=\pm0.9$ until the walls at
$z=\pm1.3$.
For numerical reasons we have assumed also a smoothing of $v_r$
between $z=\pm 0.9$ and $z=\pm 1.0$ (in DEA 3 and IEA 3),
and between
$z=\pm 0.9$ and $z=\pm 0.95$ (in DEA 4). In IEA 4, a
sharp decrease to zero
of $v_r$ at $z=\pm 0.9$ has been assumed.
The effects are astonishing. For DEA 3 and IEA 3 we get a $R^c_m$ of
94 and 97, respectively. For DEA 4 we get 128, and for IEA 4 we get
133. Apart from the fact that the rotational flow drives $R^c_m$
to forbidding high values there is also a strong sensitivity
on the precise form of
$v_r$ at the impeller position.

The second flow pattern is a rotation which is assumed constant
between
the impeller and the wall. The resulting $R^c_m$ is 98,
both for DEA 5 and IEA 5. Note that in this case
also a smoothing of $v_r$
between $z=\pm 0.9$ and $z=\pm 1.0$ was used.

\begin{figure}
\begin{center}
\begin{tabular}{c}
\epsfxsize=8.0cm\epsfbox{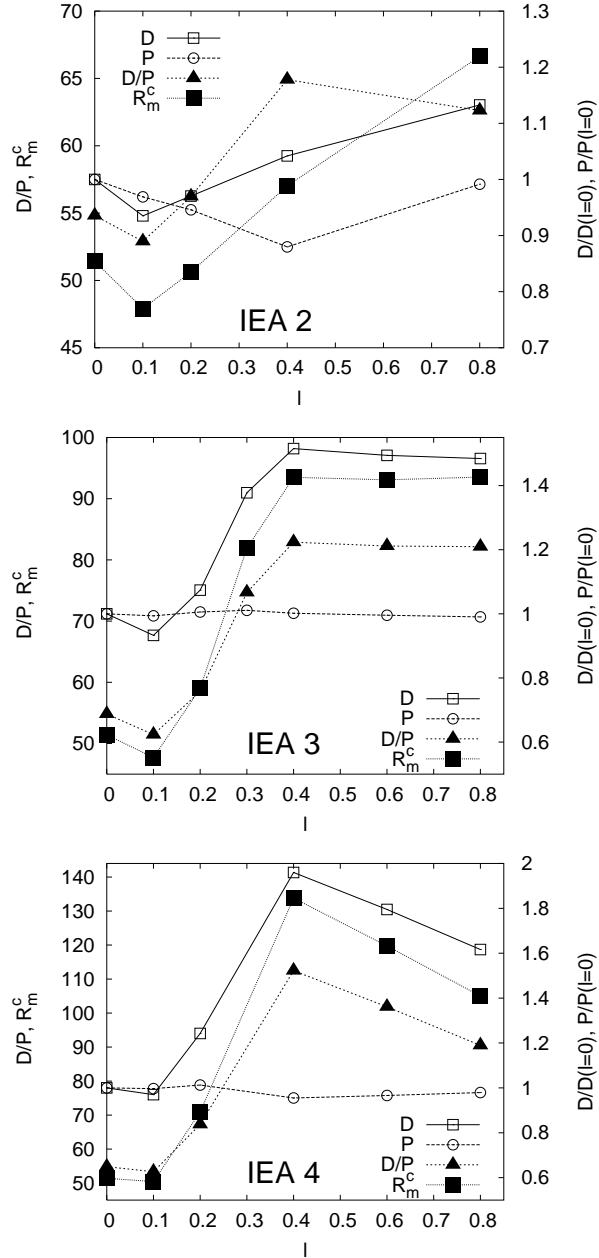}
\end{tabular}
\end{center}
\caption{Dissipation rate D, production rate P (both normalized to
the value at $l=0$),
ratio D/P, and $R^c_m$ in dependence on the
thickness $l$ of the lid layer. From the top: IEA 2, IEA 3, IEA 4.
Note the fact that the ratio D/P and $R^c_m$ are
approximately the same.}
\end{figure}

As this "lid effect" was not noticed before and may prove dangerous
for any experimental fluid dynamos driven by counter-rotating
impellers, we will try in the following to give more details on the
balance between magnetic generation and dissipation.
For this it is useful to analyze, by taking values from
IEA 2, IEA 3, and IEA 4,
a few global characteristics which are
essential for the efficiency of dynamos. From the r.h.s of
Eq. (1) we read off
that the dynamo effect is governed by
the competition of magnetic field
production (the first term) and dissipation (the second term).
Therefore we will consider in the following
the normalized dissipation rate
$D$ and the normalized
production rate $P$, defined by
\begin{eqnarray}
D&=&\frac{\int|\nabla \times {\bf{B}}|^2 \; dV}{\int |{\bf B}|^2 \; dV}\; ,\\
P&=&\frac { \int [ (\nabla \times {\bf{B}}) \times {\bf{B}}]
\cdot {\bf {u}} \; dV}{\int |{\bf B}|^2 \; dV}\; ,
\end{eqnarray}
respectively.

\begin{figure}
\begin{center}
\begin{tabular}{c}
\epsfxsize=8.0cm\epsfbox{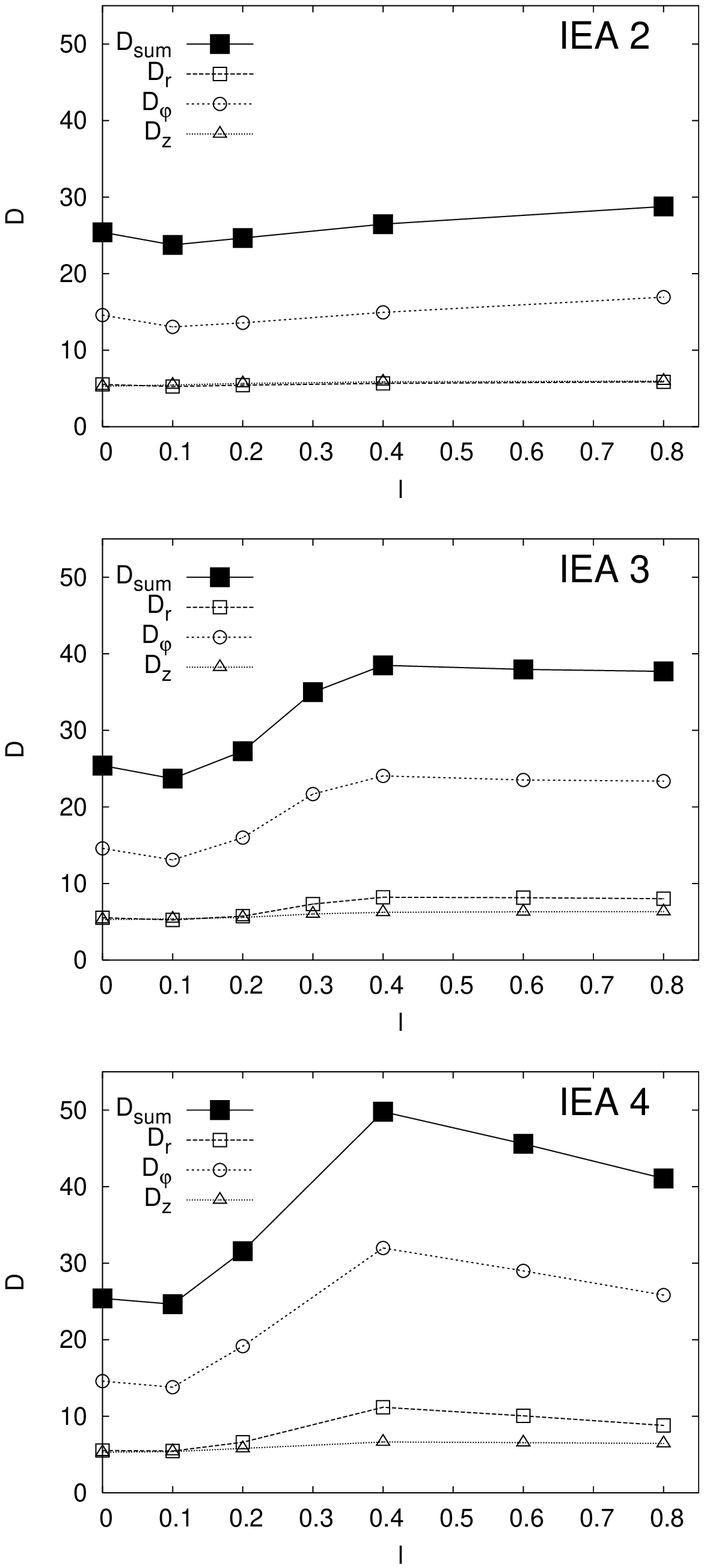}
\end{tabular}
\end{center}
\caption{Individual dissipation rates resulting from
 the radial, the azimuthal and the
axial current, and their sum. From the top: IEA 2, IEA 3, IEA 4.}
\end{figure}

\begin{figure}
\begin{center}
\begin{tabular}{c}
\epsfxsize=8.0cm\epsfbox{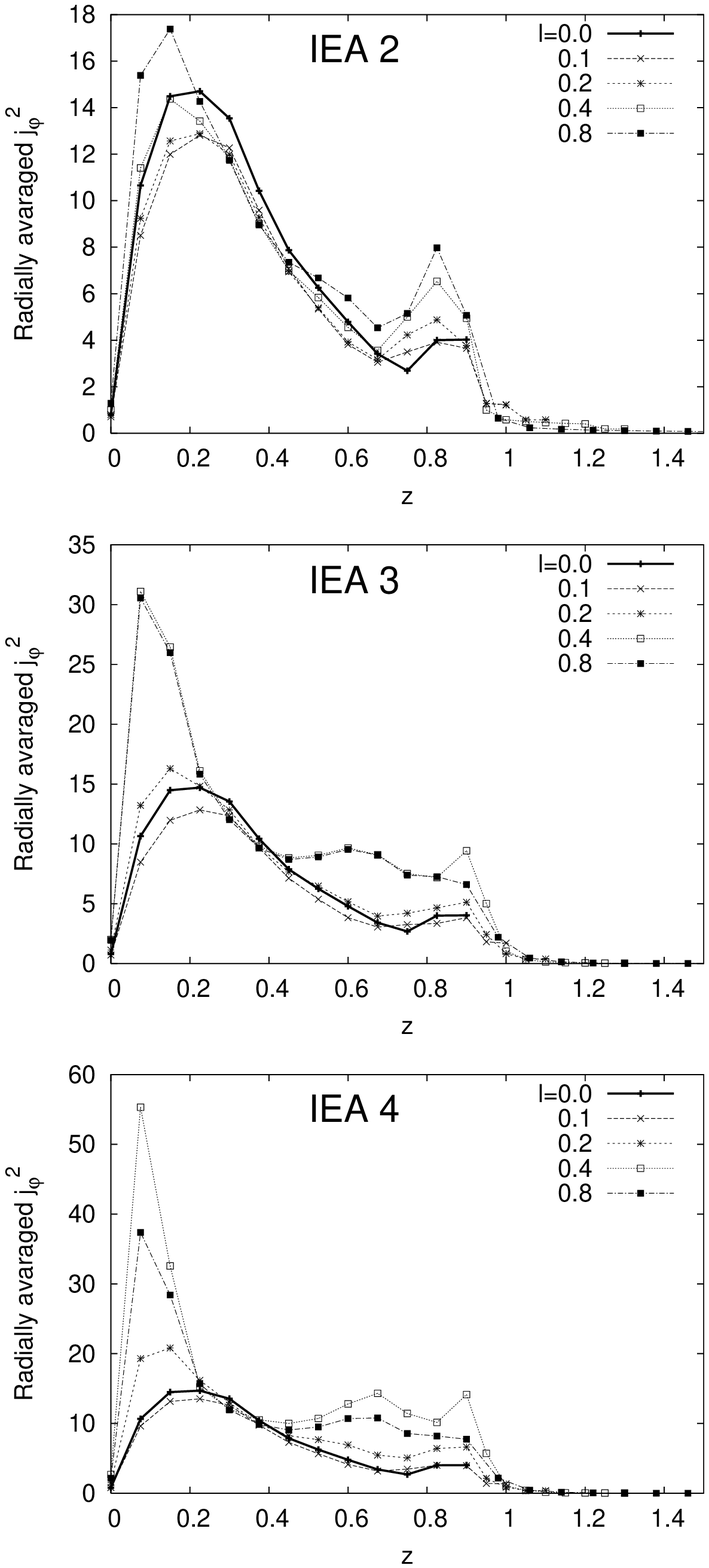}
\end{tabular}
\end{center}
\caption{$\int^{R+w}_0 j_{\varphi}^2 r dr/
(\int |{\bf B}|^2 \; dV)$ in dependence on $z$.
From the top: IEA 2, IEA 3, IEA 4. Note the different scales of the
axis of ordinates.}
\end{figure}

The dependence of these quantities, and of their ratio,
on the lid layer thickness $l$
is shown in Fig. 7, together with the corresponding $R^c_m$.

\begin{figure}
\begin{center}
\begin{tabular}{c}
\epsfxsize=8.0cm\epsfbox{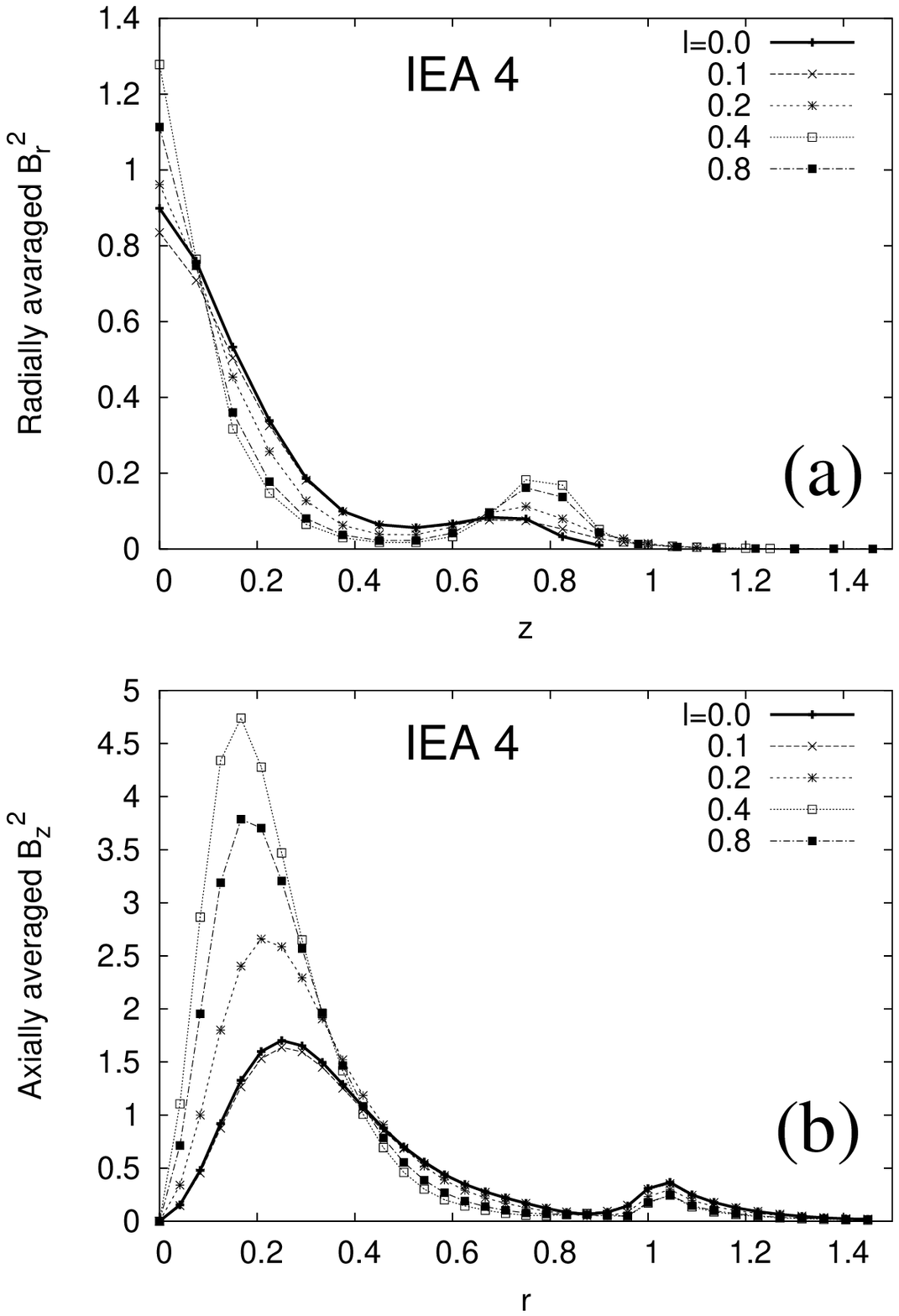}
\end{tabular}
\end{center}
\caption{Explanation of the strongly enhanced azimuthal currents
by the magnetic field components for IEA 4.
(a) $\int^{R+w}_0 B_r^2 r dr/(\int |{\bf B}|^2 \; dV)$
in dependence
on $z$ for different lid layer thicknesses $l$. Note the stronger
gradients for $l=0.4$ and $l=0.8$. (b) $\int^{H+l}_{-H-l}
B_z^2 dz/(\int |{\bf B}|^2 \; dV)$ in dependence
on $r$.}
\end{figure}

For each of the settings, IEA 2, IEA 3, and IEA 4, we observe
initially (between
$l=0$ and $l=0.1$), a decrease of the dissipation $D$, which is
not completely
compensated by a decreasing field production $P$.
As a result, we get a decrease of $D/P$ and hence of $R^c_m$.
For larger $l$, however, $D$ increases
again. In particular for IEA 3 and IEA 4 it is
interesting to see
that $P$ remains rather constant.
For IEA 4, $D$ becomes very large
for $l=0.4$
which leads to the high value of $R^c_m=133$.
Note that these global results for  production and dissipation 
strongly differ from the ones obtained with a side layer of variable 
thickness (see figure 19 in \cite{LEORAT}).

\begin{figure}
\begin{center}
\begin{tabular}{c}
\epsfxsize=11.0cm\epsfbox{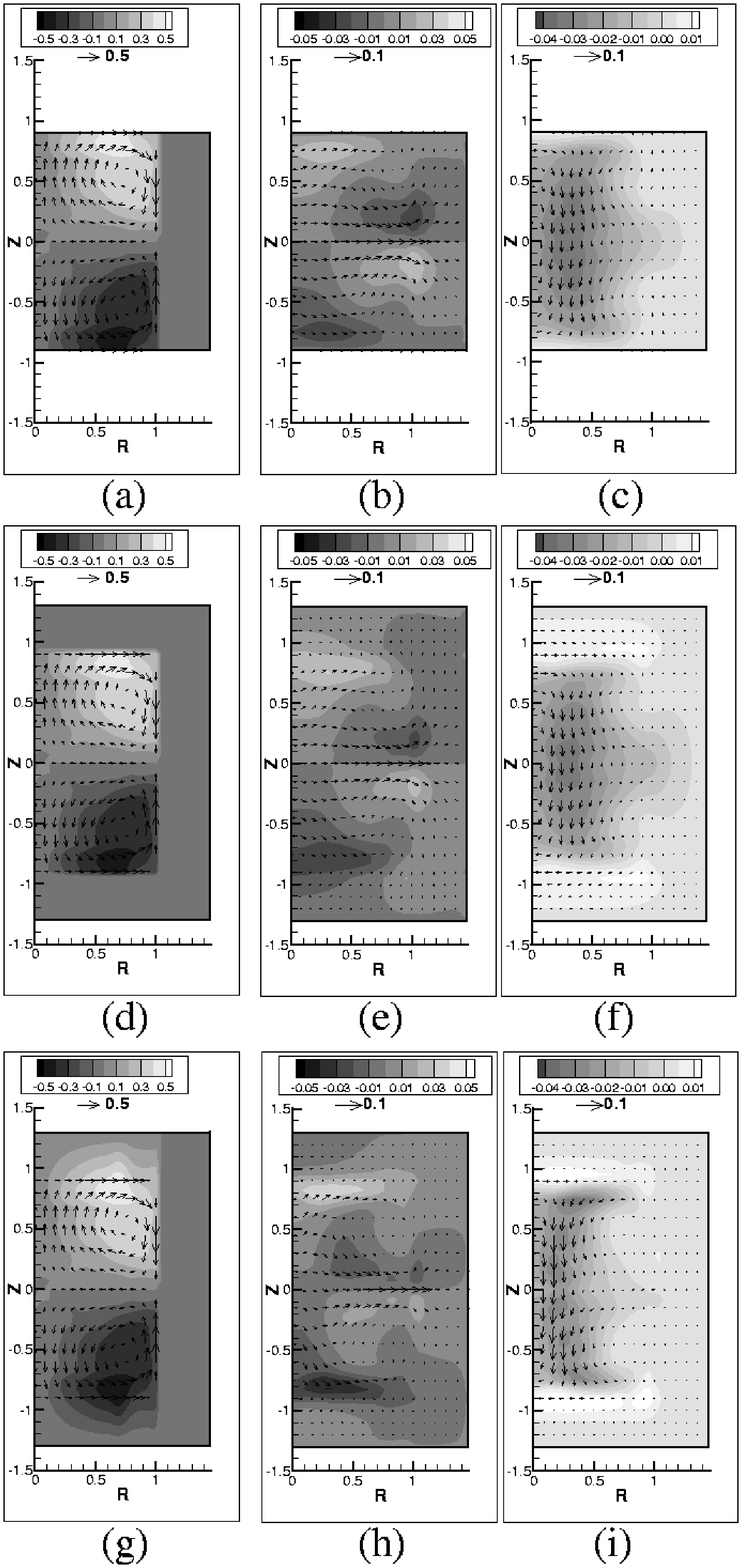}
\end{tabular}
\end{center}
\caption{Velocity field (left column)  and magnetic field structure 
in two azimuthal planes (middle and right column) 
for the VKS experiment,
assuming different
lid layer  configurations. (a,b,c) IEA 1. (d,e,f) IEA 2. (g,h,i) IEA 4.}
\end{figure}

In the following, we will analyze the impact of
lid layers on the radial, azimuthal and axial
components of the dissipation term.
We see in Fig. 8 that the radial and the axial
currents contribute relatively
little to the total dissipation, and that this contribution
does not change significantly with
increasing $l$. The dominant
dissipation, which comes from the azimuthal current,
strongly increases for IEA 3 and IEA 4 with increasing $l$.

Having identified  the azimuthal current as the ''main offender''
for the observed high $R^c_m$, let us consider it more in detail.
For this purpose,
we show in Fig. 9 the radially averaged value of $j^2_{\varphi}$,
again normalized with the total magnetic energy, in
dependence on $z$ (note that this picture is symmetric
for negative $z$).
As expected, nothing special happens for IEA 2, but for IEA 3
and even more for
IEA 4 we see a strong enhancement, in particular between
$z=0$ and $z=0.2$.

We know that $j_{\varphi}$ depends
both on the radial derivative of $B_z$ and the axial
derivative of $B_r$. For that reason,
the radial average of $B^2_r$ in dependence on $z$
and the axial average of $B^2_z$ in dependence on $r$ are depicted
in Fig. 10. We see that both fields contribute to the
enhancement of $j_{\varphi}$.

This is also visible in the magnetic field plots of Fig. 11, which
shows again two azimuthal planes. Differing from the previous
plots, we have chosen here
the settings IEA 1, IEA 2, and IEA 4. In the latter (Fig. 11f) we
see nicely how $B_z$
acquires strong radial gradients. It seems that $B_z$
is compressed into a small radial stripe, an effect that
contributes strongly
to the high azimuthal current and hence to high dissipation there.

\section{Discussion}

In this paper, we have found a strong influence of lid layers
and, even more pronounced,
of the flow structure
therein on the self-excitation threshold for von K\'arm\'an type
dynamos.
For the realistic VKS geometry, and taking into account
only static lid layers,
we obtained an increase of $R^c_m$ by around 12 per cent.
Allowing the azimuthal velocity to decay linearly
to zero between the impeller and the top and the bottom walls,
or allowing
it to be constant between the
impeller and the walls, leads to an increase of $R^c_m$ by 100
per cent to values around 95 or even by 150 percent to 133,
depending on other
flow details. Apart from accuracy questions in the order
of a few per cent, these  results should be considered as rather
reliable, since both the DEA and the IEA code yield basically the
same values and eigenfunctions.

It would be tempting to illustrate this behaviour
in simple terms.
Consider a cylinder of height $H$ and
radius $r$. Assume, for the moment, uniform currents
flowing in axial, radial and azimuthal directions.
For those, we may define individual
radial ($R_r$), azimuthal ($R_{\varphi}$), and axial
($R_z$) resistances:
\begin{eqnarray}
R_r&=&1/(2 \pi \sigma H)\; ,\\
R_{\varphi}&=&\pi/(\sigma H)\; ,\\
R_z&=&H/(\sigma \pi r^2)  \;.
\end{eqnarray}
When a stagnant {\it lid} layer ($l \ne 0$) is added at both ends
of the cylinder, the height $H$ of the cylinder becomes
larger. From Eqs. (10-12),
one can see that $R_r$ and $R_{\varphi}$ decrease and
$R_z$ increases with $H$.
For small layer thicknesses the azimuthal and radial
currents would still be dominant,
hence the whole dissipation should decrease. But if the
axial current becomes stronger, the dissipation
would increase. Under
the assumption that the production rate did not change
too much, $R^c_m$ should
show a similar behaviour.
In contrast to that, if we add
a stagnant
{\it side} layer  ($w \ne 0$),
the radius $r$ increases and the height $H$ of the cylinder
remains
unchanged.
For this situation,  from Eqs. (11-13)
one  reads off that $R_r$ and $R_{\varphi}$ would not
change, but $R_z$ would decrease. If
we assumed the production rate to
have no significant change, $R^c_m$ should always decrease.
Therefore, one
could hope that
the different
scalings of Eqs. (11-13)
with $H$ and $r$
might already reflect the
main  difference that lid layers and side layers have on the
dynamo condition.

\begin{table}
\caption{Table with different side/lid layer and flow configurations for
the VKS flow
with propeller TM 73, and the computed $R^c_m$. Note that the
experimentally achievable $R_m$ is approximately 55.
The ''side layer smoothing'', which means a linear decay to zero 
of the velocity
between $r=1$ and $r=1.1$, has no significant effect on $R^c_m$. In contrast
to this, smoothing of $v_r$ in the lid layer has drastic effects
(cp. the difference of DEA/IEA 3 and DEA/IEA 4.)}
\vspace*{0.5cm}
\begin{center}
\begin{tabular}{|p{1.8cm}|r|  p{1.5cm}|r| p{2.8cm}|p{2.8cm}|r|}
\hline
Model& $w$ &Side layer smoothing& $l$ &  $v_{\varphi}$ in lid layer
&Smoothing of $v_r$ in lid layer &$R^c_m$\\
\hline
PER 0 & 0 & - & 0 &-&-& 180\\
IEA 0 & 0 & - & 0 &-&-& 190\\
\hline
PER 1 & 0.4 & no & 0 &-&-& 43\\
DEA 1 & 0.45 & yes& 0 & -&- &48\\
IEA 1 & 0.45 & no & 0 & -&- &51\\
\hline
DEA 2 & 0.45 & no & 0.4 &no&no & 56\\
IEA 2 & 0.45 & no & 0.4 &no&no & 57\\
\hline
DEA 3 & 0.45 & yes& 0.4 & Linear decay
in $0.9<|z|<1.3$ &  Linear decay
              in $0.9<|z|<1.0$ & 94\\
IEA 3 & 0.45 & no & 0.4 &  Linear decay
in $0.9<|z|<1.3$&  Linear decay
              in $0.9<|z|<1.0$& 96\\
\hline
DEA 4 & 0.45 & yes& 0.4 &  Linear decay
in $0.9<|z|<1.3$    &  Linear decay
              in $0.9<|z|<0.95$  & 128\\
IEA 4 & 0.45 & no & 0.4 &   Linear decay
in $0.9<|z|<1.3$  &no & 133\\
\hline
DEA 5 & 0.45 & yes & 0.4 & Constant
in  $0.9<|z|<1.3$      &   Linear decay
in $0.9<|z|<1.0$  & 98\\
IEA 5 & 0.45 & no & 0.4 &  Constant
in  $0.9<|z|<1.3$       &  Linear decay
in $0.9<|z|<1.0$  & 98\\
\hline
\end{tabular}
\end{center}
\end{table}

Unfortunately, the reality seems not to be such simple.
If those scaling arguments were correct, Fig. 8a
should show a decreasing (with $l$) $D_{\varphi}$ (and $D_r$),
and an
increasing $D_z$.
In reality, however, both the minimum and
the following increase rely exclusively on the
behaviour of $D_{\varphi}$
while $D_z$ (and $D_r$) do not play any significant
role.

By following, in section 4,
the trace from enhanced dissipation via
increased azimuthal current, we have identified
the strong steepening of the axial and radial
field components
in the central bulk of the dynamo as the main reason for the
dramatic increase of $R^c_m$.
Hence, the deterioration of the dynamo condition does not rely,
as could be hypothesized, on an increased
dissipation {\it in the added lid layer}, but on the
change of the eigenmode structure
{\it in the bulk} of the dynamo.

What could be a technical consequence for the real
VKS experiment if $R^c_m$ increases significantly
due to the lid effect?
Provided that the kinematic dynamo results based on
the time-averaged flow
still apply, in essence, to
the turbulent VKS2 flow, the installed motor
power will not allow anymore
the observation of dynamo action, since the
driving power scales as
the cube of the magnetic Reynolds number. One could
thus try,
first, to
avoid any flow behind the impeller by some simple inserts or blades.
But this would lead, at the best, to an $R^c_m \sim 57$ for the
actual value $l=0.4$.
The best solution of the lid layer problem might be to install additional
steel housings between the impellers and the end walls in order
to hinder any sodium to go there \footnote{In this context 
it might be interesting 
to remind that the first dynamo of Lowes and Wilkinson became
only operative after inserting a small amount of insulation
to suppress a degenerative part of the induced currents \cite{LOWI2}.}.

Another part of the problem which has not been considered yet is the 
presence of metallic driving impellers of finite conductivity. The 
conductivity jump between the fluid and the solid has not been 
implemented in the present DEA and IEA codes (although there are no principle 
obstacles to do that). With the 
available codes, a thin layer of liquid sodium in solid rotation could 
also be an approximation of the real impeller. Since the tested rotating 
lid configurations led to threshold increase, one would generally expect 
an extra threshold increase for the complete problem. Thus, it seems 
favorable to use propellers of the smallest possible conductivity. But even 
a stainless steel impeller, which is less conducting than sodium, might still 
deteriorate the dynamo condition.
A detailed investigation of this problems is left for future work.

Various experiments using MHD flows without internal walls are now in 
progress in the world, and one expects manifestation of dynamo action 
and nonlinear saturation under conditions closer to the natural 
astrophysical dynamos.  In the case of the cylindrical VKS 
experiment, we have shown that an added layer (static or
rotating) brings about  surprizing consequences when situated in the 
lids area. These results suggest that the role of added layers be 
systematically studied for  other flow configurations, both in order 
to empirically lower the critical magnetic Reynolds number and also  
to shed some light on the process of dynamo action.

\section*{Acknowledgments}
This work was supported by
Deutsche Forschungsgemeinschaft
in frame of SFB 609 and Grant No. GE 682/12-2.
We thank the French GdR Dynamo No 2060.

\section*{Appendix}
In this appendix, we give a few indications how the 
three dimensional integral equation system (2-4) can be 
reduced to a two-dimensional one.
We consider a cylinder with the
radius $R$ and height $H$. Introducing the cylindrical coordinate 
system ($r, \varphi, z$), we have
\begin{eqnarray}
{\mathbf{r}}&=&[r \cos\varphi, r \sin\varphi, z]^T, {\mathbf{r}}'=[r' \cos\varphi', 
r' \sin\varphi',z']^T, \nonumber\\
{\mathbf{B}}&=&[B_r, B_\varphi, B_z]^T, {\mathbf{u}}=[u_r, u_\varphi, u_z]^T.
\end{eqnarray}
The magnetic field ${\mathbf{B}}$, the electric potential $\phi$, and the vector 
potential ${\mathbf{A}}$ are expanded in azimuthal modes:
\begin{eqnarray}{\label{eq4a}}
\pmatrix{{\mathbf{B}}\cr
\phi\cr
{\mathbf{A}}}=\sum^{\infty}_{m=-\infty} \pmatrix{{\mathbf{B}}_m\cr \phi_m\cr {\mathbf{A}}_m} \exp(im\varphi).
\end{eqnarray}
If the velocity field is axisymmetric, the equation system (2-4) is decoupled with 
respect to $m$. By introducing the
six integrals
\begin{eqnarray}
E_1^m(r,r',z,z')&=&\int_0^{2\pi}\frac{\cos m\varphi'}{(r^2+r'^2-2rr' 
\cos\varphi'+(z-z')^2)^{\frac{3}{2}}}d\varphi',\\
E_c^m(r,r',z,z')&=&\int_0^{2\pi}\frac{\cos m\varphi' \cos\varphi'}{(r^2+r'^2-2rr' 
\cos\varphi'+(z-z')^2)^{\frac{3}{2}}}d\varphi',\\
E_s^m(r,r',z,z')&=&\int_0^{2\pi}\frac{\sin m\varphi' \sin\varphi'}
{(r^2+r'^2-2rr' \cos\varphi'+(z-z')^2)^{\frac{3}{2}}}d\varphi',\\
D_s^m(r,r',z,z')&=&\int_0^{2\pi}\frac{\sin \varphi' \sin m\varphi'}{(r^2-2rr' \cos\varphi'+r'^2+(z-z')^2)^{\frac{1}{2}}}d\varphi' ,\\
D_c^m(r,r',z,z')&=&\int_0^{2\pi}\frac{\cos \varphi' \cos m\varphi'}{(r^2-2rr' \cos\varphi'+r'^2+(z-z')^2)^{\frac{1}{2}}}d\varphi' ,\\
D_1^m(r,r',z,z')&=&\int_0^{2\pi}\frac{\ cos m\varphi' }{(r^2-2rr' \cos\varphi'+r'^2+(z-z')^2)^{\frac{1}{2}}}d\varphi'.
\end{eqnarray}
over the angle $\varphi'$, 
all the three-dimensional integrals 
in Eqs. (2-4) can be reduced to two-dimensional 
integrals over  $r$ and $z$, and the 
two-dimensional integral can be reduced to a line integral.

Therefore, under the assumption that the velocity field is axisymmetric, the 
integral equation system (2-4) 
can be reduced to the two-dimensional case.


\begin{thebibliography}{99}
\bibitem{RUHO}G. R\"udiger and R. Hollerbach, The Magnetic Universe,
Wiley-VCH, Weinheim, 2004.
\bibitem{LOWI1} F.J. Lowes,  I. Wilkinson, Geomagnetic dynamo: a laboratory model,
Nature 198 (1963) 1158--1160.
\bibitem{LOWI2} I. Wilkinson, The contribution of laboratory 
dynamo experiments to our understanding of the
mechanism of generation of planetary magnetic fields,
Geophys. Surveys 7 (1984) 107--122.
\bibitem{RMP} A. Gailitis, O. Lielausis, E. Platacis, G. Gerbeth,
F. Stefani, Colloquium: Laboratory experiments on hydromagnetic dynamos,
Rev. Mod. Physics. 74 (2002) 973--990.
\bibitem{PRL1} A. Gailitis et al.,
Detection of a flow induced magnetic field eigenmode in the Riga dynamo facility,
Phys. Rev. Lett. 84 (2000) 4365-4368.
\bibitem{PRL2} A. Gailitis et al.,
Magnetic field saturation in the Riga dynamo experiment,
Phys. Rev. Lett. 86 (2001) 3024-3027.
\bibitem{MUST} U. M\"uller, R. Stieglitz,
Can the Earth's magnetic field be simulated in the laboratory?
Naturwissenschaften 87 (2000)
381-390.
\bibitem{STMU} R. Stieglitz, U. M\"uller,
Experimental demonstration of a homogeneous two-scale dynamo,
Phys. Fluids 13 (2001) 561-564.
\bibitem{PLASMA} A. Gailitis, O. Lielausis, E. Platacis, G. Gerbeth,
F. Stefani,
Riga dynamo experiment and its theoretical background,
Phys. Plasmas 11 (2004) 2838-2843.
\bibitem{VKS1} M. Bourgoin et al.,
Magnetohydrodynamics measurements in the von K\'{a}rm\'{a}n sodium experiment,
Phys. Fluids 14 (2002) 3046-3058.
\bibitem{VKS1.5}L. Mari{\'e} et al.,
Open questions about homogeneous fluid dynamos: the VKS experiment,
Magnetohydrodynamics 38 (2002) 163--176.
\bibitem{VKS2} F. P{\'e}tr{\'e}lis et al.,
Nonlinear magnetic induction by helical motion in a liquid sodium turbulent flow,
Phys. Rev. Lett. 90 (2003) 174501.
\bibitem{LEORAT} F. Ravelet, A. Chiffaudel, F. Daviaud, J. L\'{e}orat,
Towards an experimental von K\'{a}rm\'{a}n dynamo: numerical studies for an
optimized design, Phys. Fluids 17 (2005) 117104.
\bibitem{MBDL} L. Mari\'e, J. Burguete, F. Daviaud, J. L\'{e}orat,
Numerical study of homogeneous dynamo based on experimental
von  K\'{a}rm\'{a}n  type flows,
Eur. Phys. J. B 33 (2003) 469-485.
\bibitem{STEF} F. Stefani, G. Gerbeth, A. Gailitis,
Velocity profile optimization for the Riga dynamo experiment,
in: Alemany, A., Marty, Ph., Thibault, J.-P. (Eds.), Transfer Phenomena in Magnetohydrodynamics and
Electroconducting Flows, Kluwer, Dordrecht, 1999, pp. 31-44.
\bibitem{ASTRO} F. Stefani, G.  Gerbeth, K.-H. R\"adler,
Steady dynamos in finite domains: an integral equation approach ,
Astron. Nachr. 321 (2000) 65-73.
\bibitem{GUERMOND} J.-L. Guermond, J. L\'{e}orat, C. Nore,
A new Finite Element Method for magneto-dynamical problems:
two-dimensional results,
Eur. J. Mech. B 22 (2003)  555-579.
\bibitem {JCP} M. Xu, F. Stefani, G. Gerbeth,
The integral equation method for a steady kinematic dynamo problem,
J. Comp. Phys. 196 (2004)
102-125.
\bibitem{ISKAKOV} A.B. Iskakov, S. Descombes, E. Dormy,
An integro-differential formulation for magnetic
induction in bounded domains: boundary element-finite volume method,
J. Comp. Phys.  197 (2004) 540--554.
\bibitem{PRE} M. Xu, F. Stefani, G. Gerbeth,
Integral equation approach to time-dependent kinematic
dynamos in finite domains,
Phys. Rev. E 70 (2004) 056305.
\bibitem{BOUR} M. Bourgoin, P. Odier, J.-F. Pinton, Y. Ricard,
An iterative study of time independent induction effects in
magnetohydrodynamics, Phys. Fluids 16 (2004)  2529--2547.
\bibitem{BUGU} E.C. Bullard, D. Gubbins,
Generation of magnetic fields by fluid motions of global scale,
Geophys. Astrophys. Fluid Dyn. 8 (1977) 43-56.
\bibitem{AVALOS2} R. Avalos-Zu\~{n}iga, F. Plunian,
Influence of inner and outer walls electromagnetic
properties on the onset of a stationary dynamo,
Eur. Phys. J. B 47 (2005) 127--135.
\bibitem{TILGNER} R. Kaiser, A. Tilgner,
Kinematic dynamos surrounded by a stationary conductor,
Phys. Rev. E 60 (1999)  2949--2952.
\bibitem{AVALOS} R. Avalos-Zu\~{n}iga, F. Plunian, A. Gailitis,
Influence of electromagnetic boundary conditions
onto the onset of dynamo action in laboratory experiments,
Phys. Rev. E 68 (2003) 066307.
\bibitem{LEORAT0} J. L\'eorat, Numerical simulations of cylindrical dynamos,
AIAA Progr. Astron. Aeron. 162 (1994) 282--292.
\bibitem{MND} L. Mari\'e, C. Normand, and F. Daviaud,
Galerkin analysis of kinematic dynamos in the von K\'{a}rm\'{a}n  geometry,
Phys. Fluids 18 (2004) 017102.
\bibitem{DUJA} M.L. Dudley, R.W. James, Time-dependent
kinematic dynamos with stationary flows,
Proc. R. Soc. Lond. A 425 (1989) 407--429.
\end{thebibliography}
\end{document}